\documentclass[journal=nalefd,manuscript=article]{achemso}

	\usepackage{bm}
 	\usepackage{graphicx}
	\usepackage{hyperref}		% hyperlinks in the pdf, has to be (almost) last %
	\usepackage[figure]{hypcap}		% corrects hyperlinks to figures %
	\usepackage{amsmath}
	\usepackage{amsfonts}
	\usepackage{amssymb}
	\hypersetup{
		unicode		= false,	% non-Latin characters in Acrobat's bookmarks
		pdfborder		= {0 0 0}	% style of the border around a link; {0 0 0} gives no border %
		pdftoolbar		= true,		% show Acrobat's toolbar %
		pdfmenubar		= true,		% show Acrobat's menu %
		pdffitwindow		= false,     	% window fit to page when opened %
		pdfstartview		= {FitH},    	% fits the width of the page to the window %
		pdfnewwindow	= true,      	% links in new window %
		colorlinks		= true,		% false: boxed links; true: colored links %
		linkcolor		= blue,		% color of internal links %
		citecolor		= blue,		% color of links to bibliography %
		filecolor		= blue,		% color of file links %
		urlcolor		= blue,		% color of external links %
		linktocpage		= true,		% whole entry or only page in the toc is made into a link %
		pdftitle		= {Bilayer graphene on hexagonal substrates: Interplay between misalignment, interlayer asymmetry and trigonal warping},
		pdfauthor		= {M. Mucha-Kruczynski, J. Wallbank, V.I. Fal'ko},
		pdfsubject		= {},
		pdfcreator		= {MikTeX},
		pdfproducer		= {Marcin Mucha-Kruczynski},
		pdfkeywords		= {},
}

	\newcommand{\vect}[1]{\boldsymbol{#1}}
	
	\newcommand{\op}[1]{\hat{\boldsymbol{#1}}}

\author{M.~Mucha-Kruczy\'{n}ski}
\affiliation{Department of Physics, University of Bath, Claverton Down, Bath BA2~7AY, United Kingdom}
\email{m.mucha-kruczynski@bath.ac.uk}
\author{J.~R.~Wallbank}
\affiliation{Department of Physics, Lancaster University, Lancaster, LA1~4YB, United Kingdom}
\author{V.~I.~Fal'ko}
\affiliation{Department of Physics, Lancaster University, Lancaster, LA1~4YB, United Kingdom}

\title{Heterostructure bilayer graphene--hBN: \\ Interplay between misalignment, interlayer asymmetry, and trigonal warping}

\begin{document}

\begin{abstract}
We study the superlattice minibands produced by the interplay between moir\'{e} pattern induced by hexagonal BN substrate on graphene layer and the interlayer coupling in bilayer graphene with Bernal stacking (BLG). We compare moir\'{e} miniband features in BLG, where they are affected by the interlayer asymmetry of BLG--hBN heterostructure and trigonal warping characteristic for electrons in Bernal-stacked bilayers with those found in monolayer graphene.
\end{abstract}
%\pacs{73.22.Pr, 42.30.Ms, 81.05.ue}
\maketitle

%\section{Introduction}

The heterostructures of graphene with other hexagonal layered crystals or crystals with hexagonal symmetry facets feature moir\'{e} patterns which are the result of incommensurability of the periods of the two two-dimensional lattices, or their misalignment. Since the period of the moir\'{e} pattern is longer for a pair of crystals with a closer size of lattice constants and better aligned principal crystallographic axes, the long period moir\'{e} superlattices are characteristic for graphene/hexagonal boron nitride (hBN) heterostructures with a small misalignment angle $\theta$ between the two honeycomb lattices. Such heterostructures have recently been created by transferring graphene onto hBN \cite{xue_natmat_2011, yankowitz_natphys_2012, ponomarenko_nature_2013, hunt_science_2013}. The influence of hexagonal moir\'{e} patterns on Dirac electrons in monolayer graphene (MLG) has been studied in detail \cite{xue_natmat_2011, yankowitz_natphys_2012, ponomarenko_nature_2013, hunt_science_2013, sachs_prb_2011, kindermann_prb_2012, ortix_prb_2012, wallbank_prb_2013}, using both specific microscopic models and phenomenologically. Three possible types of moir\'{e} miniband structures on the conduction/valence band sides of graphene's spectrum have emerged from the theories \cite{yankowitz_natphys_2012, ponomarenko_nature_2013, hunt_science_2013, sachs_prb_2011, kindermann_prb_2012, ortix_prb_2012, wallbank_prb_2013}: sometimes, spectra without a distinct separation between the lowest and other minibands; quite exceptionally, the first miniband separated from the next band by a triplet of secondary Dirac points (sDPs) in each of the graphene valleys $K$ and $K'$; more generically, a single sDP at the edge of the first miniband in each valley. Also, the signatures of the miniband formation have been observed experimentally in the tunneling density of states \cite{yankowitz_natphys_2012} and magnetotransport characteristics \cite{ponomarenko_nature_2013, hunt_science_2013} in MLG/hBN heterostructures. 

\begin{figure}[t]
\centering
\includegraphics{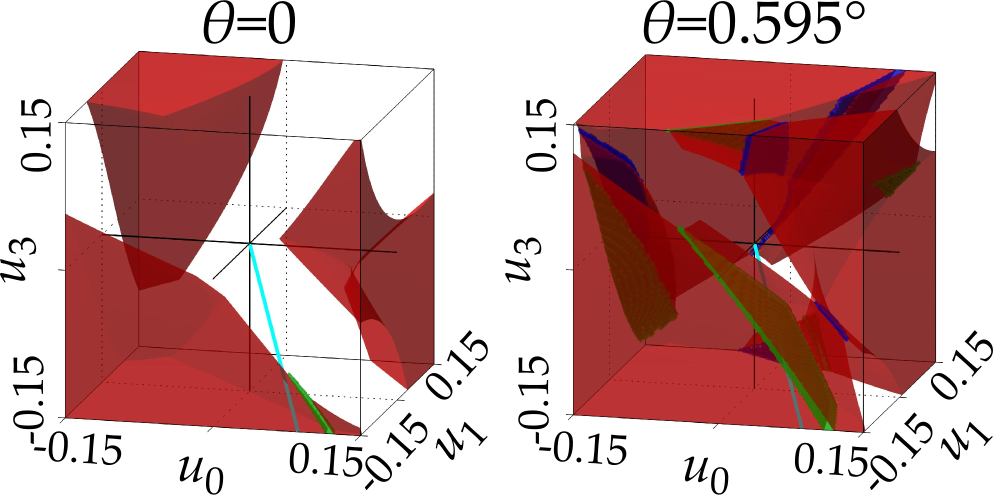}
\caption{Parameter space $(u_{0},u_{1},u_{3})$ used to classify characteristic behaviour of moire miniband in highly oriented BLG on almost commensurate substrate: in red, we paint regions where BLG spectrum has a gap separating the first miniband in the valence band from the rest of the spectrum, regions with overlapping (unresolved) bands are transparent and blue/green mark the degenerate conditions for the appearance of isolated secondary Dirac points at the first miniband edge, like in MLG \cite{yankowitz_natphys_2012, kindermann_prb_2012, ortix_prb_2012, wallbank_prb_2013}. The thick cyan lines show directions in the parameter space favoured by the point charge lattice \cite{wallbank_prb_2013} and graphene-hBN hopping \cite{kindermann_prb_2012} models. A similar parametric plot describing the minibands on the conduction band side can be obtained by inversion in the $u_{0}$-$u_{3}$ plane.}
\label{fig:diagrams}
\end{figure}

In this Letter, we analyze the characteristic moir\'{e} miniband features in heterostructures of bilayer graphene (BLG) with highly oriented and almost commensurate, hexagonal crystals, such as hBN, recently created and investigated using magneto-transport measurements by Dean et al.~\cite{dean_nature_2013}. We find that, in contrast to monolayers, the electronic spectrum of BLG on hBN is most likely to exhibit gaps between the first moir\'{e} miniband and the rest of the spectrum (on valence or conduction band side, and sometimes in both bands), or have the bands strongly overlapping with each other, whereas Dirac points at the miniband edge appear only for exceptionally unique choice of moir\'{e} parameters. Also, we find that a gap at the edge between the valence and conduction bands can be opened in BLG by the same moir\'{e} perturbation that would not open a `zero-energy' gap in MLG. This behaviour is prescribed by the substrate creating a moir\'{e} perturbation only for one layer of BLG, thus breaking the inversion symmetry of the moir\'{e} superlattice. The results of a systematic study of the miniband regimes in BLG--hBN heterostructures is summarized in the parametric space diagrams in Fig.~\ref{fig:diagrams}, where the regions of the parameter space with gapped spectra are painted in red with an overlapping (non-resolved) bands - left transparent. Differences between the two diagrams corresponding to different misalignment angles $\theta$ arise from the interplay between the orientation of the supercell Brillouin zone (sBZ) and the skew interlayer hopping in BLG. This interplay, unique to BLG, can help in narrowing down the microscopic parameters of moir\'{e} pattern at graphene--hBN interface using magnetotransport and capacitance experiments.

		%%%%% Fig - diagrams %%%%%%%%
		%%%%% End of Fig - diagrams %%%%%

\begin{figure}[t]
\centering
\includegraphics{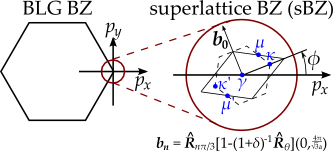}
\caption{Hexagonal Brillouin zone (BZ) of BLG with the valley coordinate system $p_{x},p_{y}$ and the electronic bands in the vicinity of the $K$ valley, together with the zoom in on the supercell Brillouin zone (sBZ) where we mark its own symmetry points $\kappa$ and $\mu$.}
\label{fig:bzs}
\end{figure}

The analysis in this paper is performed using a phenomenological approach \cite{wallbank_prb_2013} which involves the description of the long-range moir\'{e} superlattice using a Dirac-type model for graphene electrons, where we include all symmetry-allowed terms in the moir\'{e} perturbation applied to one of the two layers in BLG and perform an exhaustive numerical analysis in order to characterise the miniband behaviour over a broad range of the parameter space. For BLG placed on top of a substrate with hexagonal symmetry and the lattice constant $a_{\mathrm{S}}\!\!=\!\!(1\!+\!\delta)a$ larger by $\delta$ than that of graphene, ($a\!\!=\!\!2.46$\AA, and for the case of hBN, $\delta\!\!=\!\!1.8\%$ \cite{xue_natmat_2011}), the lattice mismatch, together with a possible misalignment of the two lattices given by the angle $\theta$, lead to a periodic structure which can be described using a set of reciprocal lattice vectors,
\begin{align}
\vect{b_{n}}\!\!=\!\!\op{R}_{n\pi/3}\!\left[ 1\!-\!(1\!+\!\delta)^{-1}\op{R}_{\theta} \right]\!(0,\frac{4\pi}{\sqrt{3}a}), \,\,\,\,\, n\!=\!0,1,\dots,6, \nonumber
\end{align}
where $\op{R}_{\varphi}$ stands for anticlockwise rotation by angle $\varphi$ and $b\!=\!\!|\vect{b_{n}}|\!\!\approx\!\!\frac{4\pi}{\sqrt{3}a}\!\sqrt{\delta^{2}\!+\!\theta^{2}}$ \cite{hunt_science_2013}. Note that this set both rotates by $\phi(\theta)$ and changes its size as a function of $\theta$, Fig.~\ref{fig:bzs}. Because of the rapidly decaying nature of the interlayer interaction, we only take into account the influence of hBN on the bottom carbon layer, neglecting any interaction with the top one. Then, following from the MLG investigations \cite{wallbank_prb_2013}, the electrons in BLG/hBN heterostructure are described by the Hamiltonian 
\begin{align}\label{eqn:4x4}
& \!\!\!\!\!\! \op{H} \!\!\! =\!\!\!\! \left(\!\!\!\!\! \begin{array}{cc} v\vect{\sigma}\!\cdot\!\vect{p}\!+\!\delta\!\op{H}_{\mathrm{sym}}\!+\!\delta\!\op{H}_{\mathrm{asym}} & \!\!\!\!\frac{1}{2}\gamma_{1}\!(\tau_{z}\sigma_{x}\!-\!i\sigma_{y})\!+\!\frac{1}{2}v_{3}\!(\sigma_{x}\!+\!i\tau_{z}\sigma_{y})\!(p_{x}\!+\!i\tau_{z}p_{y}) \\ \frac{1}{2}\gamma_{1}\!(\tau_{z}\sigma_{x}\!+\!i\sigma_{y})\!+\!\frac{1}{2}v_{3}\!(\sigma_{x}\!-\!i\tau_{z}\sigma_{y})\!(p_{x}\!-\!i\tau_{z}p_{y}) & \!\!\!\!v\vect{\sigma}\!\cdot\!\vect{p} \end{array} \!\!\!\!\!\right)\!\!\!, \\
& \!\!\!\!\!\! \delta\!\op{H}_{\mathrm{sym}} = u_{0}vbf_{1}(\vect{r}) \!+\! u_{3}vbf_{2}(\vect{r})\sigma_{z}\tau_{z} \!+\! u_{1}v[\vect{l_{z}}\!\times\!\nabla f_{2}(\vect{r})]\!\cdot\!\vect{\sigma}\tau_{z} \!+\! u_{2}v\nabla f_{2}(\vect{r})\!\cdot\!\vect{\sigma}\tau_{z}, \nonumber \\
& \!\!\!\!\!\! \delta\!\op{H}_{\mathrm{asym}} = \tilde{u}_{0}vbf_{2}(\vect{r}) \!+\! \tilde{u}_{3}vbf_{1}(\vect{r})\sigma_{z}\tau_{z} \!+\! \tilde{u}_{1}v[\vect{l_{z}}\!\times\!\nabla f_{1}(\vect{r})]\!\cdot\!\vect{\sigma}\tau_{z} \!+\! \tilde{u}_{2}v\nabla f_{1}(\vect{r})\!\cdot\!\vect{\sigma}\tau_{z}, \nonumber \\
& \!\!\!\!\!\! f_{1}(\vect{r}) = \sum_{n}e^{i\vect{b_{n}}\cdot\vect{r}}, \,\,\,\,\,\,\, f_{2}(\vect{r}) = i\sum_{n}(-1)^{n}e^{i\vect{b_{n}}\cdot\vect{r}}, \nonumber
\end{align}
written in the basis of the Bloch states on sublattices $\{\phi(A_{1}),\phi(B_{1}),\phi(A_{2}),\phi(B_{2})\}$ in the $K$ valley and $\{\phi(B_{1}),-\phi(A_{1}),\phi(B_{2}),-\phi(A_{2})\}$ in $K'$, where indices 1/2 mark the bottom/top layers and the substrate directly acts on the electrons in the bottom layer. We also use $\hbar\!=\!1$ and employ two sets of Pauli matrices $\sigma_{i}$, $\vect{\sigma}\!=\!(\sigma_{x},\sigma_{y})$, and $\tau_{i}$, acting in the sublattice and valley space, respectively. 

		%%%%% Fig - BZs %%%%%%%%
		%%%%% End of Fig - BZs %%%%%

The intralayer Dirac-like terms $v\vect{\sigma}\!\cdot\!\vect{p}$ on the diagonal and interlayer off-diagonal terms on the right-hand side of Eq.~(1) follow from the tight-binding description of BLG \cite{mccann_prl_2006}, with $v\!=\!10^{6}$m/s resulting from the interlayer hopping in graphene. Also, $\gamma_{1}\!=\!0.38$eV \cite{kuzmenko_prb_2009}, characterizes the `vertical' interlayer hopping between the closest neighbours and $v_{3}\approx 0.12v$\cite{mccann_prl_2006, kuzmenko_prb_2009, kechedzhi_prl_2007} described the trigonal warping of BLG bands, resulting from the skew next-neighbour interlayer hopping. This results in the four-band BLG model, with two bands degenerate at the points $K$ and $K'$, and two bands split away by $\pm\gamma_{1}$.

The moir\'{e} perturbation, scaled using energy scale $vb$ and parametrised using dimensionless $\{u_{i},\tilde{u}_{i}\}$, $i\!=\!0,1,2,3$\cite{wallbank_prb_2013}, captures the effect of the substrate on BLG through its coupling with carbon orbitals in the closest, bottom layer only. Following the approach used in the monolayer study \cite{wallbank_prb_2013}, we separate the moir\'{e} perturbation due to hBN into inversion-symmetric and asymmetric teerms, $\delta\!\op{H}_{\mathrm{sym}}$ and $\delta\!\op{H}_{\mathrm{asym}}$, respectively. By considering two limiting cases in which electrons in graphene are dominantly affected by only one of the two atoms in hBN unit cell, we argue that $|\tilde{u}_{i}|\!\ll\!|u_{i}|$ and neglect $\delta\!\op{H}_{\mathrm{asym}}$ in all further discussions of the band structures, except the analysis of a gap opened at zero energy. The remaining parameters, $u_{0,1,2,3}$, can be associated with the following characteristics of the moir\'{e} pattern: parameter $u_{0}$ characterizes the magnitude of a smooth electrostatic potential, $u_{3}$ captures the local asymmetry between the $A_{1}$ and $B_{1}$ sublattices, and $u_{1}$ and $u_{2}$ introduce modulation of the in-plane hops for the electrons travelling within the bottom layer. As opposed to the monolayer case \cite{wallbank_prb_2013}, in BLG the rotorless $u_{2}$ term cannot be completely gauged away. However, it vanishes for zero misalignment angle, and its effect on the band structure is generated via the interplay with the trigonal warping term: therefore, it is small and can be neglected. Note that, since we scale all energies by $vb\!\approx\!v\frac{4\pi}{\sqrt{3}a}\!\sqrt{\delta^{2}\!+\!\theta^{2}}$, the size of dimensionless parameters $u_{i}$ in Eq.~(1) would be larger for smaller angle $\theta$ for the same pair of BLG and a substrate. 

Due to the time-inversion symmetry, described by the operation \cite{aleiner_prl_2006, lemonik_prb_2012}
\begin{align}
\op{H}(\vect{p}) \!=\! \sigma_{y}\tau_{y}\!\left[ \op{H}^{*}(-\vect{p}) \right]\!\!\sigma_{y}\tau_{y}, \nonumber
\end{align}
electronic spectra in the two valleys are related, $\epsilon_{\vect{K}\!+\vect{p}}\!=\!\epsilon_{\vect{K^{'}}\!-\vect{p}}$, so that we only discuss electronic spectra in one valley, e.g. $K$. We find that in contrast to unperturbed BLG, the spectrum resulted from a generic choice of parameters in Hamiltonian (1) is not electron-hole symmetric, but obeys the following relation,
\begin{align}
\epsilon^{u_{0},u_{1},u_{3}}_{\vect{K}+\vect{p}} = -\epsilon^{-u_{0},u_{1},-u_{3}}_{\vect{K}+\vect{p}}, \nonumber
\end{align}
which folds the parameter space $(u_{0},u_{1},u_{3})$ to be explored.

		%%%%% Fig - band structures %%%%%%%%
\begin{figure}[t]
\centering
\includegraphics{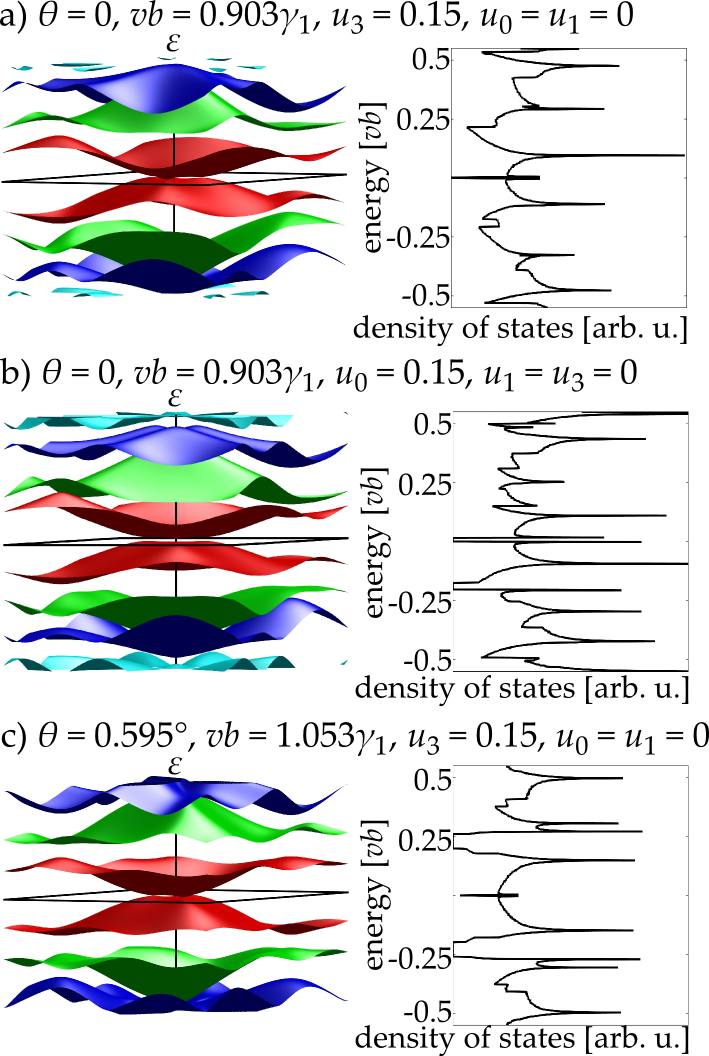}
\caption{(a-c) Moir\'{e} miniband spectra (drawn within the rhombic sBZ) and density of states (DoS) portraying two characteristic behaviours of the miniband spectrum as determined in Fig. \ref{fig:diagrams}.}
\label{fig:band_structures}
\end{figure}
		%%%%% End of Fig - band structures %%%%%

For the sake of a systematic comparison with the low-energy `two-band' model of free-standing BLG \cite{mccann_prl_2006}, applicable at the energy scale $\epsilon\!\ll\!\gamma_{1}$, we use a Schrieffer-Wolff transformation \cite{schrieffer_physrev_1966} and project the four-band Hamiltonian onto the low-energy bands, reducing it to an effective two-band Hamiltonian,
\begin{align}\label{eqn:2x2}
& \op{H}_{\mathrm{eff}} \!\approx\! -\frac{v^{2}}{\gamma_{1}}\!\left[ (p_{x}^{2}\!-\!p_{y}^{2})\sigma_{x}\!+\!2p_{x}p_{y}\sigma_{y} \right]\!\tau_{z} + v_{3}(\vect{\sigma}\!\cdot\!\vect{p})^{T} +\\ & \! \frac{vb}{2}g_{\!+}\!(\!\vect{r}\!)(1\!+\!\sigma_{z}\tau_{z}) \!+\! \frac{v^{3}b}{2\gamma_{1}^{2}}\!(p_{x}\!\!-\!\!i\sigma_{z}p_{y})g_{\!-}\!(\!\vect{r}\!)(p_{x}\!\!+\!\!i\sigma_{z}p_{y})\!(1\!\!-\!\!\sigma_{z}\tau_{z}) + \nonumber \\ & \! \frac{v^{2}b}{2\gamma_{1}}\!\left[ (p_{x}\!\!+\!\!ip_{y}\tau_{z})g(\!\vect{r}\!)(\sigma_{x}\!\!-\!\!i\sigma_{y}\tau_{z}) \!+\! g^{\!*}\!(\!\vect{r}\!)(p_{x}\!\!-\!\!ip_{y}\tau_{z})(\sigma_{x}\!\!+\!\!i\sigma_{y}\tau_{z}) \right]\!, \nonumber
\end{align} 
where
\begin{align}
& g_{\pm}(\vect{r}) \!=\!  (u_{0}\pm\tilde{u}_{3})f_{1}(\vect{r})\pm (u_{3}\pm\tilde{u}_{0})f_{2}(\vect{r}), \nonumber \\ 
& g(\vect{r}) \!=\!  \sum_{n}e^{i\vect{b_{n}}\cdot\vect{r}}(\hat{b}_{n}^{x}\!+\!i\hat{b}_{n}^{y}\tau_{z})\left[(-1)^{n}(u_{2}\!+\!iu_{1}\tau_{z})+(\tilde{u}_{1}\tau_{z}\!-\!i\tilde{u}_{2})\right], \,\,\,\,\,\,\vect{b_{n}}=b(\hat{b}_{n}^{x},\hat{b}_{n}^{y}). \nonumber
\end{align}
The applicability of the simplified Hamiltonian to the description of, at least, the first moir\'{e} miniband in the BLG spectrum requires that $\gamma_{1}\!\gtrsim\!2vb$. For a perfectly aligned BLG/hBN heterosystem, we estimate that $\frac{\gamma_{1}}{vb}\!\approx\!\frac{\sqrt{3}(1\!+\!\delta)}{4\pi}\frac{a\gamma_{1}}{v}\!=\!1.107$, which suggests that a quantitative description of moir\'{e} minibands in BLG/hBN requires the use of the four-band Hamiltonian, Eq.~(1).

Little is known about the exact values of each of the perturbation parameters $u_{i}$. Due to the size of the moir\'{e} and the importance of Van der Waals interaction in graphene/hBN heterostructures, resources needed for supercell-size ab initio calculations are prohibitive. At the same time, even for the more studied system of MLG/hBN, experimental data obtained so far also does not allow for clear determination of the perturbation \cite{ponomarenko_nature_2013}. Hence, in our numerical modelling of spectra, we broadly cover the $(u_{0},u_{1},u_{3})$ parameter space, Fig.~\ref{fig:diagrams}. Examples of characteristic moir\'{e} miniband spectra on the valence band side in BLG/hBN are shown in Fig.~\ref{fig:band_structures}: (a) overlapping minibands characteristic for the transparent part of the parameter space in Fig.~\ref{fig:diagrams}; (b) and (c) gapped spectrum at the edge of the first moir\'{e} miniband characteristic for the red-painted part of the parameter space in Fig.~\ref{fig:diagrams}. Also shown are the corresponding densities of states, with a  global gap in the valence band, Fig.~\ref{fig:band_structures}(b), (c).

		%%%%% Fig - band structures %%%%%%%%
\begin{figure}[t]
\centering
\includegraphics{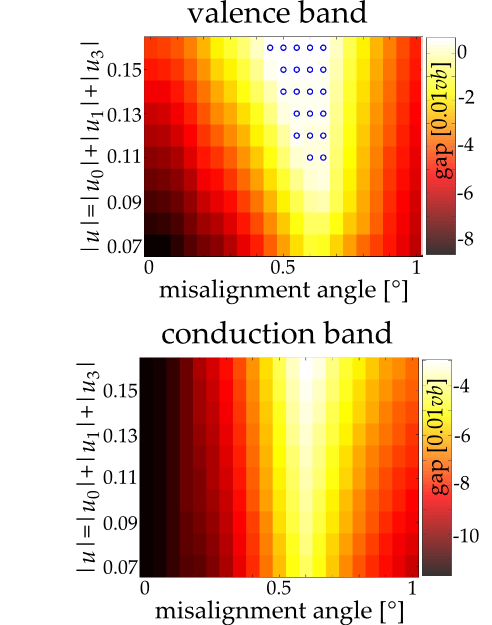}
\caption{The size of the gap between the first and second minibands on the conduction (top) and valence (bottom) side as a function of the parturbation magnitude $u\!=\!\sum_{i}|u_{i}|$ and the misalignment angle $\theta$, assuming that the perturbation parameters are described by relations in Eq.~(3). Negative values for the gap denote overlapping bands. Points marked with blue circles correspond to spectra for which the global gap is direct at $\kappa'$.}
\label{fig:angle_diagrams}
\end{figure}
		%%%%% End of Fig - band structures %%%%%

\begin{figure}[t]
\centering
\includegraphics{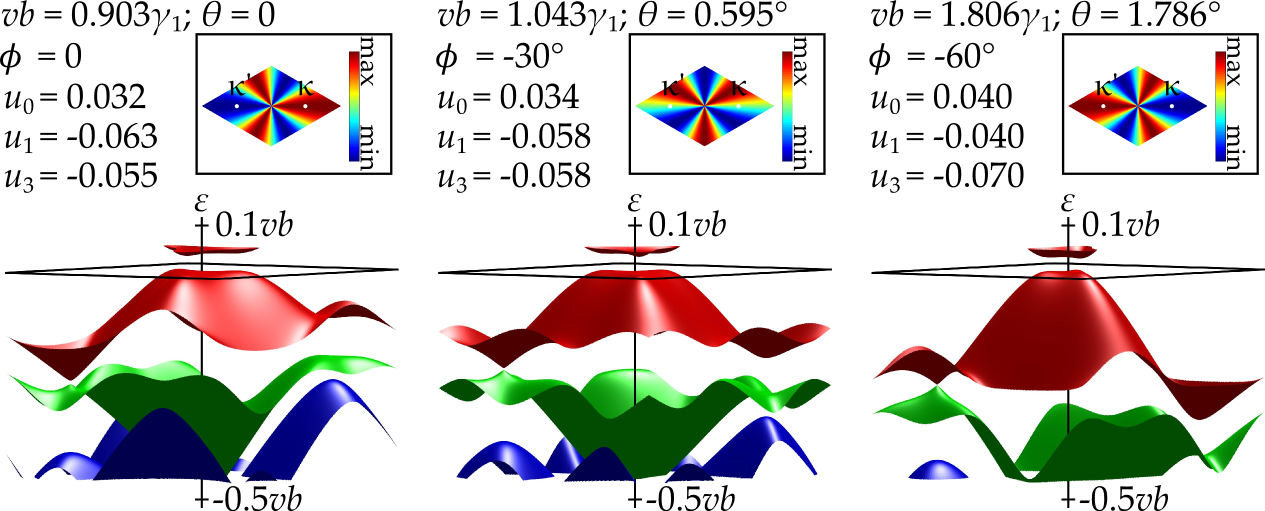}
\caption{Moir\'{e} miniband spectra (drawn within the rhombic sBZ) illustrating the role of trigonal warping in determining whether the first and second minibands on the valence side are overlapping or gapped at a secondary Dirac point. For these examples, we assumed that relations in Eq.~(3) hold and set $u\!=\!\sum_{i}|u_{i}|=0.15$. The insets show the angular dependence, $\cos\!3\phi$, of the trigonal warping for the valence band within the (rotated) sBZ.}
\label{fig:valence_band_structures}
\end{figure}

Spectra in Fig.~\ref{fig:band_structures}(a) and (c) both correspond to the same choice of perturbation, $u_{0}=u_{1}=0$, $u_{3}=0.15$, but a different misalignment angle $\theta$, which is enough to cause opening of a band gap between the first and second minibands on the valence side. This is because of the trigonal warping of the unperturbed BLG spectrum, which shifts the energy $\epsilon_{\vect{p}}^{0}$ of a momentum state $\vect{p}\!=\!(p_{x},p_{y})$ by $\epsilon_{\mathrm{warp}}\!\approx\!-sv_{3}p\cos\!3\varphi$, where $\varphi=\arctan\frac{p_{y}}{p_{x}}$, $p\!=\!\sqrt{p_{x}^{2}\!+\!p_{y}^{2}}$ and $s\!=\!1$ ($s\!=\!-1$) denotes the conduction (valence) band. For $\theta\!=\!\phi\!=\!0$, this results for example in the $\kappa$ ($\kappa'$) point in the valence band shifted up (down) in energy, like in Fig.~\ref{fig:band_structures}(a). However, misalignment angle $\theta\!=\!0.595^{\circ}$ leads to sBZ rotation by $\phi\!=\!-30^{\circ}$ and the trigonal warping correction at the points $\kappa$ and $\kappa'$ of the rotated sBZ vanishes, as seen in Fig.~\ref{fig:band_structures}(c). Such a strong dependence of the miniband spectrum on the misalignment angle is special to BLG, because in MLG trigonal warping corrections are much weaker. Consequently, experimental investigation of the BLG miniband spectra for several misalignment angles (determined from the moir\'{e} geometry) may yield new information about the nature of the perturbation felt by graphene electrons due to hBN. For example, two of the models suggested for the MLG/hBN heterostructure, graphene-hBN hopping model \cite{kindermann_prb_2012} and point charge lattice model \cite{wallbank_prb_2013},  yield the same form of the $u_{i}$ coefficients as a function of the misalignment angle $\theta$,
\begin{align}\label{eqn:us}
u_{0}=\frac{1}{2}\tilde{v}, \,\,\,\,\, u_{1}=-\frac{\delta}{\sqrt{\delta^{2}+\theta^{2}}}\tilde{v}, \,\,\,\,\, u_{3}=-\frac{\sqrt{3}}{2}\tilde{v},
\end{align} 
($\tilde{v}\!>\!0$ is a strength of the perturbation), corresponding to a single line for each misalignment angle $\theta$, as shown with bold cyan lines in Fig.~\ref{fig:diagrams} for $\theta\!=\!0$ and $\theta\!=\!0.595^{\circ}$. It is interesting to note that in the diagram for $\theta\!=\!0.595^{\circ}$, this line passes close behind the green region where a secondary Dirac point at $\kappa'\!=\!(-\frac{1}{\sqrt{3}}b,0)$ separates the first and the second miniband on the valence side. In Fig.~\ref{fig:angle_diagrams}, we show a detailed study of the gap between the first and second miniband in the conduction band as a function of the misalignment angle $\theta$ and perturbation magnitude $u\!=\!\sum_{i}|u_{i}|$, assuming that Eq.~(3) holds, where `negative' gap means that the minibands overlap. However, for the valence band, due to trigonal warping, a gap opens for $\theta\!\approx\!0.6^{\circ}$; for points marked with a circle, this is a direct gap at the $\kappa'$ point. Note that even a small change of misalignment angle leads to a large rotation of moir\'{e} pattern and a different involvement of trigonal warping due to skew interlayer coupling in BLG. The corresponding evolution of BLG moir\'{e} spectra is illustrated in Fig.~\ref{fig:valence_band_structures}. For $\theta\!=\!0$, left column in Fig.~\ref{fig:valence_band_structures}, trigonal warping affects $\kappa$ and $\kappa'$ in the opposite fashion (shown in the inset) what obscures the sDP at $\kappa'$. However, the sDP becomes isolated on the energy scale for $\theta\!\approx\!0.5^{\circ}$, centre column in Fig.~\ref{fig:valence_band_structures}. It disappears again as the misalignment angle is increased further, with the asymmetry between $\kappa$ and $\kappa'$ reaching maximum for $\theta\!=\!1.786^{\circ}$ when the sBZ is rotated by $60^{\circ}$, right column in Fig.~\ref{fig:valence_band_structures} \cite{footnote}. 

		%%%%% Fig - band structures %%%%%%%%
		%%%%% End of Fig - band structures %%%%%

One can notice that in majority of the spectra presented in Fig.~\ref{fig:band_structures} and \ref{fig:valence_band_structures}, and for generic moir\'{e} perturbation, a gap, $\Delta$, is opened at the `neutrality point' between the conduction and valence bands, in contrast to monolayer graphene, where such gap,
\begin{align}\label{eqn:gap_mlg}
\Delta^{\mathrm{MLG}}\!=\!24vb(u_{1}\tilde{u}_{0}\!+\!u_{0}\tilde{u}_{1}),
\end{align}
appears only when moir\'{e} pattern contains an inversion asymmetric perturbation. In the monolayer, this is accompanied by an overall shift, $12vb(u_{1}u_{3}\!+\!\tilde{u}_{1}\tilde{u}_{3})$, resulting in the edges of conduction/valence band at $\epsilon^{\mathrm{MLG}}_{\pm}\!=\!12vb(u_{1}u_{3}\!+\!\tilde{u}_{1}\tilde{u}_{3})\!\pm\!12vb(u_{1}\tilde{u}_{0}\!+\!u_{0}\tilde{u}_{1})$. In BLG, one of the two degenerate zero-energy states belongs to the bottom and one to the top layer. Because the top layer is unaffected by the perturbation, the corresponding state stays at zero energy, while the other one is shifted by $\epsilon^{\mathrm{MLG}}_{+}$, just like one of the states in MLG. As a result, the bilayer gap is nonzero even if the moir\'{e} perturbation is inversion-symmetric,
\begin{align}\label{eqn:gap_blg}
\Delta^{\mathrm{BLG}}\!\approx\!12vb|u_{1}\!(u_{3}\!+\!\tilde{u}_{0})\!+\!\tilde{u}_{1}\!(u_{0}\!+\!\tilde{u}_{3})|.
\end{align}
as a consequence of the hBN substrate breaking the equivalence of the two layers and, hence, the inversion symmetry. For BLG sandwiched between two hBN layers, the asymmetry will be still present, if the misalignment angles between graphene and top/bottom hBN layers are different.

In summary, we showed that the interplay between interlayer coupling (including skew hopping between layers leading to the trigonal warping effect in BLG) in bilayer graphene and breaking of layer symmetry by the substrate play an important role in determining the miniband spectrum of BLG--hBN heterostructures. As opposed to MLG in which a gap at the Dirac point is open only for inversion-asymmetric moir\'{e} perturbation, in BLG a `zero-energy' gap is open even for an inversion-symmetric perturbation, as a direct consequence of interlayer asymmetry caused by the substrate.

%\begin{acknowledgments}
This work has been supported by the ERC Advanced Grant {\it Graphene and Beyond}, EU STREP {\it ConceptGraphene}, CDT NOWNano, Royal Society Wolfson Research Merit Award and EPSRC Science and Innovation Award.

%\end{acknowledgments}


\begin{thebibliography}{99}

\bibitem{xue_natmat_2011} Xue J.; Sanchez-Yamagishi J.; Bulmash D.; Jacquod P.; Deshpande A.; Watanabe K.; Taniguchi T.;, Jarillo-Herrero P.; LeRoy B.~J.~\href{http://dx.doi.org/10.1038/nmat2968}{{\it Nat.~Mat.}~{\bf 10}, 282 (2011)}.
% STM of MLG/hBN

\bibitem{yankowitz_natphys_2012} Yankowitz M.; Xue J.; Cormode D.; Sanchez-Yamagishi J.~D.; Watanabe K.; Taniguchi T.; Jarillo-Herrero P.; Jacquod P.; LeRoy B.~J.~\href{http://dx.doi.org/10.1038/nphys2272}{{ \it Nat.~Phys.}~{\bf 8}, 382 (2012)}.
% MLG superlattices - experiment and theory containing u0

\bibitem{ponomarenko_nature_2013} Ponomarenko L.~A.; Gorbachev R.~V.; Yu G.~L.; Elias D.~C.; Jalil R.; Patel A.~A.; Mishchenko A.; Mayorov A.~S.; Woods C.~R.; Wallbank J.; Mucha-Kruczynski M.; Piot B.~A.; Potemski M.; Grigorieva I.~V.; Novoselov K.~S.; Guinea F.; Fal'ko V.~I.; Geim A.~K. \href{http://dx.doi.org/10.1038/nature12187}{{\it Nature} {\bf 497}, 594 (2013)}.
% Hofstadter's butterflies in MLG/hBN

\bibitem{hunt_science_2013} Hunt B.; Sanchez-Yamagishi J.~D.; Young A.~F.; Yankowitz M.; LeRoy B.~J.; Watanabe K.; Taniguchi T.; Moon P.; Koshino M.; Jarillo-Herrero P.; Ashoori R.~C. \href{http://dx.doi.org/10.1126/science.1237240 }{{\it Science} {\bf 340}, 6139 (2013)}.

\bibitem{sachs_prb_2011} Sachs B.; Wehling T.~O.; Katsnelson M.~I.; Lichtenstein A.~I. \href{http://link.aps.org/doi/10.1103/PhysRevB.84.195414}{{\it Phys.~Rev.~B} {\bf 84}, 195414 (2011)}.
% MLG/hBN DFT

\bibitem{kindermann_prb_2012} Kindermann M.; Uchoa B.; Miller D.~L. \href{http://link.aps.org/doi/10.1103/PhysRevB.86.115415}{{\it Phys.~Rev.~B} {\bf 86}, 115415 (2012)}.
% theory of graphene on hBN

\bibitem{ortix_prb_2012} Ortix C.; Yang L.; van den Brink J. \href{http://link.aps.org/doi/10.1103/PhysRevB.86.081405}{{\it Phys.~Rev.~B} {\bf 86}, 081405 (2012)}.
% theory of graphene on hBN - only diagonal terms

\bibitem{wallbank_prb_2013} Wallbank J.~R.; Patel A.~A.; Mucha-Kruczynski M.; Geim A.~K.; Fal'ko V.~I. \href{http://link.aps.org/doi/10.1103/PhysRevB.87.245408}{{\it Phys.~Rev.~B} {\bf 87}, 245408 (2013)}.
% symmetry-based analysis of MLG/hBN

\bibitem{dean_nature_2013} Dean C.~R.; Wang L.; Maher P.; Forsythe C.; Ghahari F.; Gao Y.; Katoch J.; Ishigami M.; Moon P.; Koshino M.; Taniguchi T.; Watanabe K.; Shepard K.~L.; Hone J.; Kim P. \href{http://dx.doi.org/10.1038/nature12186}{{\it Nature} {\bf 497}, 598 (2013)}.
% Hofstadter's butterflies in BLG/hBN

\bibitem{mccann_prl_2006} McCann E.; Fal'ko V.~I. \href{http://link.aps.org/doi/10.1103/PhysRevLett.96.086805}{{\it Phys.~Rev.~Lett.}~{\bf 96}, 086805 (2006)}.
% BLG Hamiltonian

\bibitem{kuzmenko_prb_2009} Kuzmenko A.~B.; Crassee I.; van der Marel D.; Blake P.; Novoselov K.~S. \href{http://link.aps.org/doi/10.1103/PhysRevB.80.165406}{{\it Phys.~Rev.~B} {\bf 80}, 165406 (2009)}.
% BLG tight-binding parameters

\bibitem{kechedzhi_prl_2007} Kechedzhi K.; Fal'ko V.~I.; McCann E.; Altshuler B.~L. \href{http://link.aps.org/doi/10.1103/PhysRevLett.98.176806}{{\it Phys.~Rev.~Lett.}~ {\bf 98}, 176806 (2007)}.
% BLG antilocalization

\bibitem{lemonik_prb_2012} Lemonik Y.; Aleiner I.~L.; Fal'ko V.~I. \href{http://link.aps.org/doi/10.1103/PhysRevB.85.245451}{{\it Phys.~Rev.~B} {\bf 85}, 245451 (2012)}.

\bibitem{aleiner_prl_2006} Aleiner I.~L.; Efetov K.~B. \href{http://link.aps.org/doi/10.1103/PhysRevLett.97.236801}{{\it Phys.~Rev.~Lett.}~{\bf 97}, 236801 (2006)}.
% time-reversal symmetry in monolayer graphene

\bibitem{schrieffer_physrev_1966} Schrieffer J.~R.; Wolff P.~A. \href{http://link.aps.org/doi/10.1103/PhysRev.149.491}{{\it Phys.~Rev.}~{\bf 149}, 491 (1966)}.
% Schrieffer-Wolff transformation

\bibitem{footnote} We tested all of our above conclusions for a "single-side gate" geometry, where large charge density required to fill the first miniband (four electrons per moir\'{e} supercell) is induced by a single gate, which induces additional interlayer asymmetry \cite{mccann_prb_2006, mucha-kruczynski_ssc_2009}, and found that its presence has almost no effect on the type of the miniband spectrum at the edge of the first miniband (gapped or overlapping minibands).

\bibitem{mccann_prb_2006} McCann E. \href{http://link.aps.org/doi/10.1103/PhysRevB.74.161403}{{\it Phys.~Rev.~B} {\bf 74}, 161403 (2006)}.

\bibitem{mucha-kruczynski_ssc_2009} Mucha-Kruczynski M.; McCann E.; Fal'ko V.~I. \href{    http://dx.doi.org/10.1016/j.ssc.2009.02.057}{{\it Sol.~St.~Commun.}~{\bf 149}, 1111 (2009)}.
% self-consistent calculation of the gap in bilayer graphene

\end{thebibliography}
\end{document}